\documentstyle[iopconf1]{article}
\begin{document}
\title{Is Dark Matter made up of
\\Massive Quark Objects?\footnote{Presented
by S.F. at Dark 98, Heidelberg, July 20-24, 1998;
to be published in the Proceedings}}

\author{Sverker Fredriksson\dag,
Daniel Enstr\"{o}m\dag,
Johan Hansson\dag,
Svante Ekelin\ddag
\ and Argyris Nicolaidis\S}

\affil{\dag\ Department of Physics,
Lule\aa \ University of Technology, \\
SE-971 87 Lule\aa , Sweden}

\affil{\ddag\ Department of Mathematics,
Royal Institute of Technology,
SE-100 44 Stockholm, Sweden}

\affil{\S\ Department of Theoretical Physics,
Aristotle University of Thessaloniki,
GR-540 06 Thessaloniki, Greece}

\beginabstract
We suggest that dark matter is made up of
massive quark objects that have survived
from the Big Bang, representing the ground state of
``baryonic'' matter. Hence, there was no
overall phase transition of the original
quark matter, but only a split-up
into smaller objects. We speculate that
normal hadronic matter comes about
through enforced phase transitions when
such objects merge or collide, which also gives
rise to the cosmic gamma-ray bursts.
\endabstract

\section{Introduction}
In this talk I will try to convince you that dark matter consists
of objects made up of quarks in the form of so-called
quark-gluon plasma (QGP), with up to a few solar masses.
This model builds on an idea by
Witten from 1984 \cite{Witten84}, but is more radical in the sense
that our quark objects are not just ``nuggets'',
nor do just ``contribute'' to dark matter. We also
suggest that cosmic gamma-ray bursts (GRBs) result
from mergers of quark objects, which
relates dark matter to GRBs in a unique way.
A more detailed account of the model can
be found in \cite{Hansson98}.

Witten's idea was
much discussed a decade ago,
but seems out of fashion today, which might be
due to the current interest in
supersymmetry, neutralinos and the like.
The notion of ``QGP in space'' is still popular, but
mainly in models for neutron stars,
or other high-density objects \cite{Glendenning96}.

We believe that time is ripe for reviving
quarks as the source of dark matter. The main motivation
is one of {\em simplicity}. It is generally acknowledged that all
normal matter was once upon
a time a quark-gluon plasma. One
way or the other, it went through a phase transition,
and turned into our world. What would be more natural than
to assume that something went ``wrong'', so that most
of this plasma remained as dark matter?
It seems less natural to invent a third form of
matter (axions, neutralinos,...),
unrelated to our matter, and without
experimental support.

Since our dark-matter model
requires quark objects to be less energetic than
normal hadronic matter, {\em i.e.}, representing the
absolute ground-state of matter, I will start by discussing
the stability aspects.

\section{The Stability of Quark Matter}

It was suggested in 1971 by Bodmer
\cite{Bodmer71} that a QGP might represent the
ground state of matter. This idea was
strengthened by many
analyses \cite{Chin79,Bjorken79,Witten84,
Rujula84,Fahri84}. It became clear
that such objects could be ``dark'',
because the best ground-state candidate is a QGP
consisting of equal (or almost equal) parts of $u$,
$d$ and $s$ quarks, making it (almost) electrically
neutral.

These ideas cannot be rigorously proven from
first principles, {\em i.e.}, from QCD.
Rather, one has to rely on
phenomenological models, and
the one most frequently used for analysing
quark objects is the so-called
MIT bag model \cite{Chodos74,DeGrand75},
where quarks are confined to ``bags'' due to an
external ``vacuum'' pressure, quantified by the
so-called bag constant
$B$ (quoted as $B^{1/4} \approx 150$ MeV).
Many versions of the MIT bag
model have been developed, containing various
corrections, one of which is for interactions among
the quarks, which were assumed to be free inside the
bag in the original version. A recent review of
strange quark matter is given in \cite{Greiner98}.

The original MIT bag model results
for protons and other low-mass hadrons were
achieved through exact solutions of the Dirac equation
inside a spherical bag. Various approximations
must be applied for systems with many quarks.
Typically, systems with more than
a few dozen quarks are lighter than the corresponding
atomic nuclei \cite{Greiner98}. This result is not
violated by the apparent stability of heavy nuclei, since
it is utterly improbable that they would decay into
``strangelets'' of equally many $u$, $d$ and $s$ quarks.

For extremely heavy objects gravity
plays an important role, and the MIT bag
model (or any other microscopic model)
must be complemented with general
relativity. This is normally done by applying the
so-called Tolman-Oppenheimer-Volkoff (TOV)
equation \cite{TOV39} from 1939, which describes
a static, spherical, ideal-liquid system, where
the stability is equivalent to
a zero-pressure at the surface of the object.
It is derived in many textbooks in general
relativity, {\em e.g.}, \cite{Schutz90}.
Several such analyses of astrophysical systems
can be found in the literature. They are based on
different equations of state (with or without
quark effects), and other detailed
assumptions.

Some recent computations by
one of us \cite{Enstrom97} give results that are
typical for analyses in the spirit of the MIT
bag model (or other bag models).
A QCD-based equation of state for
a system of (equally many) $u$, $d$ and $s$ quarks
was used.
As an example, the
quark objects have masses below around
$\sim 1.8M_{\odot}$ ($M_{\odot}$ being the solar
mass) and radii below $\sim 11$ km when $B^{1/4} = 150$ MeV.
Similar values can be found in analyses of
the stability of neutron stars, in particular those
with a sizeable quark-matter core \cite{Glendenning96}.

Here one might ask how one and the same approach
can lead to predictions of a quark core inside
neutron stars, as well as of the existence of pure quark stars.
Should not one of these systems be the absolute ground-state
of matter? The answer might be that a neutron star is created
by contraction of normal nuclear matter (with $u$ and $d$
quarks) in a supernova, while ``our'' quark stars are created
from the primordial quark gluon plasma with equal numbers
of the three lightest quarks. Therefore, the final states need
not be identical. A decay of a neutron (hybrid) star into
a pure quark star could be utterly improbable.

\section{A Novel Big Bang Scenario}

In Witten's model \cite{Witten84}, the early
universal plasma went through a phase
transition into hadronic matter. However, thanks to the
high pressure of the hadronic gas, a fraction of the
QGP managed to survive in the form of quark nuggets.
Their high internal pressure was balanced by the
outer hadronic pressure until they cooled down
and survived on their own (being the
ground state of matter).

It is unrealistic to assume that such a scenario could lead
to a final state where $90-99$ {\em per cent}
would be quark objects. We therefore suggest that there
was no overall
phase transition at all! The present Universe is instead
a result of an early {\em split-up} of the plasma into
smaller pieces of all possible sizes. We assume that
the internal forces inside the original QGP
could not withstand the rapid expansion, and the plasma
``cracked'' along surfaces of lower density, or weakened
quark forces. This is, in fact, a normal chain of events
for classical explosions.

The further development depends on the scale
of this first split-up, {\em i.e.}, the mass distribution
among the first generation of quark objects, which seems
impossible to estimate from first principles.
It could be due to events during the
inflation phase ({\em e.g.}, quantum fluctuations), or to
random occurrence of colour-neutral subregions.
Neither is it possible to apply arguments built on
the concept of an event horizon
\cite{Uggla98}.
We do not know when the split-up took place. It could
have been when the plasma was strongly super-cooled,
{\em i.e.}, when it had a density lower than that of nuclear
matter. Also, the concept of a horizon loses its meaning
inside a high-density, uniform Universe,
where it is even impossible to define a length.

We will therefore discuss three possible ways that the QGP
could have split up. Each ones has its own phenomenological
motivation.

The least interesting version is that the plasma split up directly
into $1-10$ {\em per cent} normal nucleons and
$90-99$ {\em per cent} quark objects
of up to solar masses. Or alternatively,
that nucleons quickly ``boiled off'' the surfaces of quark
objects that had not yet stabilised. This would explain
the ``observed'' fraction of normal matter, without introducing
any dramatic, subsequent events. One only needs to assume that
the further fate of the Universe was dominated by gravity.

However, this scenario does not explain why the normal
matter gathered mostly in the centre of galaxies, while
dark matter seems to be more peripheral. Neither
does it give a clue to the well-known problem of galaxy
formation, {\em i.e.}, how galaxies could form in less than
a billion years.

It is therefore tempting to speculate that the mass scale of
galaxies played some role during the first split-up of the plasma.
Hence, either the
original plasma separated immediately into subregions
with typical galactic masses, or there was a more or less
{\em fractal} split-up, where the galactic scale was only one
of many mass-scales. The latter alternative fits well the
findings of a fractal structure of the present Universe.

Let us start by discussing quark objects of galactic masses,
or more generally, of masses in excess of the limit given
by the Schwarzschild radius. As an example, the Milky Way,
assuming a $90$ {\em per cent} dark-matter fraction,
would have been a quark object with a radius of
around $80,000$ km (if it was spherical), and
containing some $10^{69}$ quarks.

Obviously, such objects must have been highly unstable,
and would seemingly have disappeared quickly into black holes.
However, during the very early phase their interiors were still
following the general expansion of the Universe, and they most
probably also rotated (as they do today). Consequently, it
might have happened that only the inner part of a
proto-galaxy went into a black hole, while the outer parts
continued to expand. This would explain why many
galaxies appear to have massive central black holes of masses
$(10^{6}-10^{10})M_{\odot}$ \cite{Franceschini98}.

This opens up for an interesting history of a galaxy. The
black hole would suddenly ``empty'' the inner part of the
quark object, and temporarily turn it into a hollow system.
Since there would be no resistance from gravity, nor from
a bag pressure, matter would continue to flow into the
central cavity, and probably in small-enough pieces to
condense into normal hadronic matter, while giving off radiation.
The visible galaxy would then be created as an accretion disk
(or sphere) around the central black hole.
This process ends when the radiation
pressure blows the rest of the galaxy apart.
If all visible matter in the Milky Way was created
in this process, it would have given off some
$10^{63}-10^{64}$ erg of radiation (assuming an energy gain
of $10-100$ MeV per hadron during the condensation).
This is enough to break apart the dark-matter galactic
shell into smaller quark objects, and send them out
in the distant periphery. After this violent phase, gravity
took over and formed the Milky Way of today.
Interestingly enough, it was noted already in
1958 \cite{Ambartsumian58} that some galaxies seem
to have been formed by explosions, and this idea
cannot easily be dismissed even in the light of
more recent observations \cite{Harwit88}.

Let us now turn to the less dramatic option that
the original QGP split up into pieces with masses
below the one given by gravitational stability,
{\em i.e.}, from isolated hadrons up to objects of a few
solar masses. A hint of the mass distribution
is given by the observation of machos in the periphery
of the Milky Way. These seem to have
a mass range of $(0.01-0.8) M_{\odot}$
\cite{Alcock96}, with
a mean macho mass
of roughly $0.5M_{\odot}$, and a $50$ {\em per cent}
macho fraction in the Milky Way (dark) halo
(assuming a spherical shape). Such estimates
naturally depend strongly on the assumed
total amount of dark matter. For the sake of our
model it can nevertheless be concluded that the
mass distribution is biased toward
heavy objects - from ``planetary'' to
solar masses.

Still, we assume that some
early mechanism, such as a fractal break-up
of the original plasma, made these quark objects
appear at once in proto-galactic clusters, with a
higher density in the centre than in the periphery.
If so, the next step in the development
of a galaxy should have been {\em mergers},
or collisions, of quark objects. This phase started
as soon as the interior of a proto-galaxy
was shielded from the general expansion of the
Universe, so that gravity and/or random motion
became the dominant
factors. This could have started almost ``immediately'',
{\em i.e.}, within a fraction of a second after
the first break-up of the QGP.

A merger of a binary system of quark objects
should be a dramatic event. In systems
corresponding to solutions to the TOV equation,
with a vacuum pressure typical for the MIT bag
model, the overall density is about the same as
inside a free nucleon. However, this density depends
very strongly on the pressure, and would drop
far below that of normal nuclear matter, would
the bag constant drop from the normal
$B^{1/4} = 150$ MeV to, say, $75$ MeV
\cite{Enstrom97}. This effect is strongest
at the periphery of the objects. A similar,
but weaker depletion of the density
is expected if gravity would be ``turned off''.
Both these phenomena are expected when two
massive quark objects merge,
while counterbalancing each other's internal
gravity, and screening each other from
the vacuum pressure.

All in all, we expect bridges or jets of
quark matter to appear between the two objects
just before they merge into one. This corresponds
to a more or less continuous stream of ``boiled-off''
quarks, which would hence hadronise into
normal matter, while radiating the excess energy
in the form of gammas, mesons etc. This
zone of newborn hadrons and radiation would,
in fact, be a mini-copy of the conventional
Big Bang scenario for the whole Universe.
The main difference is that the conventional
Universe found its true ground state when
expanding and cooling, while our quark
objects are forced into an ``unnatural''
hadronisation with the help of the
gravitational energy released in the binary system.
Such matter-flows occur in many cosmic
situations, {\em e.g.}, between galaxies, 
between normal stars in a binary system, and
from an accretion disc into a black hole.
There are rather well-established formalisms
for analysing such situations \cite{Carroll96}.

It is worth noting that a most
popular class of models for cosmic
gamma-ray bursts assume that they originate from
neutron-star mergers \cite{Blinnikov84}.
A majority of these models rely on some
either unspecified, or very complicated,
mechanism for converting
the energy-gain into gammas. However,
there are also models were the gamma-ray
bursts originate from a hadron-quark phase
transition inside a neutron star, or in
connection to a merger (see, {\em e.g.},
\cite{Cheng96}).
These transitions are {\em from} a hadron
phase {\em to} a quark phase, and hence
quite opposite to our scenario.

So, in this version of our model, the
early proto-galaxy will experience
a myriad of such mergers, where the most
violent ones will be between quark objects
of a few solar masses. They will result in miniversions
of the conventional overall phase transition
from quark to hadron matter, and hence
lead to normal matter mainly in the galactic
centre, and to gamma-ray bursts.

The expected
energy-release in the form of gammas is in line
with observations. Supposing that a maximal
amount of one solar mass can convert to hadrons
in a merger, this energy will be $10^{53}-10^{54}$ erg
(again assuming an energy-gain of $10-100$ MeV
per produced hadron), which fits well the
estimated energies of those bursts that have
so far been localised with the help of observed
redshifts.

As the radiated gammas cannot
penetrate more than around $100$ fm inside
dense hadron (or quark) matter, we expect the radiation
to blow the hadronic zone apart, and partly
thermalise, and hence prevent the hadrons
from falling into the quark objects again.
The quick merging of many binaries
probably led to an overall expansion of the
whole proto-galaxy due to the collective
radiation pressure.

After this violent phase, the hadrons (mostly
baryons) went through a normal development,
{\em e.g.}, nucleosynthesis into light elements,
formation of atoms etc. It has been suggested
recently \cite{Kainulainen98} that some of the
paradoxes with the fraction of light elements
might be solved if nucleosynthesis took
place in rather small, disjoint systems, of down to
kilometre sizes, which at least suggests that
the Universe split up into smaller objects
before hadronising.

In the present Universe, and in the visible
galaxies, gamma-ray bursts are rare, because
the original binary systems merged very early,
and new ones are created only by accident in the very
diluted dark halo. In this respect, our model
differs from the ones built on neutron star mergers,
because neutron stars are continuously created in
all galaxies, while quark matter is not. Hence, we
predict that gamma-ray bursts are much more
frequent in very distant galaxies than in the nearby
ones. This is in line with observations, at least of
the ``localised'' GRBs.

Also, we can only observe the most
violent ones. Maybe there are more frequent,
but invisible, mergers between quark objects of much smaller
masses. It would be interesting to analyse the observed
gamma-ray ``background'' in our own galaxy
in the light of this idea, and try to restrict some model
parameters. In any case, we expect a full range of
gamma-ray energies, contrary to models built on
neutron star mergers, because neutron stars are
more uniform in size. Hence, gamma ray bursts
are not expected to be ``standard candles''.

Finally, some of you might want to know if
we expect an alarming rate of these objects in our
own solar system. Well, they would certainly {\em not}
make up ten times the mass of the sun, because there
is a lot of empty space inbetween the stars.
So, even if our galaxy has quark objects of a total
mass of some 1 trillion solar masses, they would be
distributed in a volume much larger than that of the
visible Milky Way, and probably more or less evenly
in space (perhaps even spherically around the galaxy).
In a volume corresponding to a sphere inside
the orbit of the earth around the sun we expect
no more quark matter
than contained in a centimetre-sized
object. Nevertheless,
it would have the mass of a medium-sized mountain
(a few billion tonnes),
and it would not be of much help to call for
Bruce Willis, because there would be no
way to blow it apart,
even if we would observe it in space!

\section*{Acknowledgements}
SF would like to thank the Organisers for
arranging an excellent conference in the
beautiful city of Heidelberg, and the
Max-Planck-Institute f\"ur Kernphysik,
and in particular Professors B. Povh
and B. Kopeliovich for their hospitality
during his two-month stay in Heidelberg.
DE would like to thank Professors
J. Silk, C. Uggla and E. Witten for inspiring
discussions and constructive comments.
This project is supported by the
European Commission under contract
CHRX-CT94-0450, within the network
"The Fundamental Structure of Matter".


\vspace{-14pt}
\begin{thebibliography}{9}

\bibitem{Witten84}Witten E 1984 \PR {\rm D}{\bf 30} 272

\bibitem{Hansson98}Hansson J 1998
{\em The Fundamental Structure of Matter}
Lule\aa ~ University of Technology
PhD thesis 1998:04 CIV; see also
Enstr\"{o}m D {\em et al.} astro-ph/9802236

\bibitem{Glendenning96}Glendenning N K 1996
{\em Compact Stars} (New York: Springer-Verlag)

\bibitem{Bodmer71}Bodmer A R 1971
\PR {\rm D}{\bf 4} 1601

\bibitem{Chin79}Chin S A and Kerman A K 1979
\PRL {\bf 43} 1292

\bibitem{Bjorken79}Bjorken J D and McLerran L D 1979
\PR {\rm D}{\bf 20} 2353

\bibitem{Rujula84}De~R\'ujula A and Glashow S L 1984
Nature {\bf 312} 734

\bibitem{Fahri84}Fahri E and Jaffe R L 1984
\PR {\rm D}{\bf 30} 2379

\bibitem{Chodos74}Chodos A {\em et al.} 1974
\PR {\rm D}{\bf 9} 3471

\bibitem{DeGrand75}DeGrand T {\em et al.} 1975
\PR {\rm D}{\bf 12} 2060

\bibitem{Greiner98}Greiner C and
Schaffner-Bielich J nucl-th/9801062

\bibitem{TOV39}Oppenheimer J R and Volkoff G 1939
\PR {\bf 55} 377

\bibitem{Schutz90}Schutz B 1990 {\em A First Course
in General Relativity} (Cambridge University Press)

\bibitem{Enstrom97}Enstr\"{o}m D 1997
{\em Astrophysical Aspects of Quark-Gluon
Plasma} Lule\aa ~ University of Technology
MSc diploma thesis 1997:366 CIV, also
available as hep-ph/9802337

\bibitem{Uggla98}Uggla C 1998 private
communication

\bibitem{Franceschini98}Franceschini A, Vercellone S
and Fabian A C 1998 astro-ph/9801129

\bibitem{Ambartsumian58}Ambartsumian V 1958
report at the 1958 Solvay Conference, Brussels

\bibitem{Harwit88}Harwit M 1988
{\em Astrophysical Concepts} (New York: Springer-Verlag)

\bibitem{Alcock96}Alcock C {\em et al.} 1996
Astrophys. J. {\bf 471} 774; {\bf 486} 697 (1997)

\bibitem{Carroll96}Carroll B W and Ostlie D A 1996
{\em Modern Astrophysics} (Addison-Wesley);
see also New K C B and Tohline J E gr-qc/9703013

\bibitem{Blinnikov84}Blinnikov S I {\em et al.}
1984 SvAL {\bf 10} 177

\bibitem{Cheng96}Cheng K S and Dai Z G 1996
\PRL {\bf 77} 1210

\bibitem{Kainulainen98}Kainulainen K, Kurki-Suonio H and
Sihvola E 1998 astro-ph/9807098

\end{thebibliography}
\end{document}